\documentclass[fleqn,10pt]{wlscirep}

\usepackage{newunicodechar}
\newunicodechar{μ}{\ensuremath{\mu}}
\usepackage[T1]{fontenc}

\usepackage[utf8]{inputenc}
\usepackage[numbers,sort,square]{natbib}

\usepackage{hyperref}
\usepackage{graphicx}
\usepackage{subcaption}
\usepackage{lipsum}
\usepackage{array}
\usepackage[utf8]{inputenc}
\usepackage{fontenc}
\DeclareUnicodeCharacter{2005}{ }
\usepackage{makecell}
\usepackage{graphicx} 
\usepackage{booktabs} 
\usepackage{makecell} 

\title{High-power monolithic narrow-linewidth 1.6 mJ/8 ns fiber laser system based on all-glass spun tapered double-clad fiber amplifier}

\author[1,*]{Hossein Fathi}
\author[1]{Uttam Samanta}
\author[2]{Evgenii Motorin}
\author[2]{Ebrahim Aghayari}
\author[3]{Andrey Grishchenko}
\author[4]{Matias Koivurova}
\author[1,2]{Regina Gumenyuk}
\author[2]{Valery Filippov}

\affil[1]{Laboratory of Photonics, Physics Unit, Faculty of Engineering and Natural Sciences, Tampere University, Korkeakoulunkatu 3, 33720 Tampere, Finland}
\affil[2]{Ampliconyx Ltd, Lautakatonkatu 18, 33580, Tampere, Finland}
\affil[3]{Ceramoptec SIA, Domes street 1a, Livani, LV-5316, Latvia}
\affil[4]{Center for Photonics Sciences, University of Eastern Finland, P.O. Box 111, 80101, Joensuu, Finland}

\affil[*]{hossein.fathi@tuni.fi}


\begin{abstract}

High-energy, narrow-linewidth nanosecond pulses are highly demanding for many applications that require high temporal and spatial coherence. However, the amplification of narrow-linewidth pulses is primarily limited by stimulated Brillouin scattering, which causes pulse instabilities, back-reflected pulses, and catastrophic damage effects on optical components. In this work, we present a 1.6 mJ narrow-linewidth nanosecond pulsed fiber laser system based on all-glass
spun tapered double-clad fibers without employing any mitigating technique for the stimulated Brillouin scattering effect. The system delivers pulses with an 8~ns duration at a 100~kHz repetition rate, over 97.5\% degree of polarization, a beam quality factor of \( M^2 \sim 1.3 \), a spectral linewidth of 53.8~MHz, a 160 W average power, and 188 kW peak power with a slope efficiency of 97.6\%. The degree of spatial coherence of the amplified signal was measured to be 0.94. Our results are highly valued in applications requiring high-energy, high-coherence pulses with spectral, spatial, and polarization characteristics in a compact system.
  
\end{abstract}
\begin{document}

\flushbottom
\maketitle
\thispagestyle{empty}

\section*{Introduction}
All-fiber coherent nanosecond master oscillator power amplifier (MOPA) systems with high energy and peak power have gained significant interest over the past decade for applications such as coherent LIDAR~\cite{Philippov:04, lidar, lidar2}, deep-space communication~\cite{deep-space}, the generation of visible and UV light for lithography~\cite{fourth-harmonic,Tsubakimoto:17,775nm}, as well as coherent beam combining and spectral beam combining~\cite{CBC-narrow, Fathi2021,Schmidt:09}.
These applications typically demand highly coherent optical pulses (linewidths of a few tens of MHz), relatively short pulses of several nanoseconds, peak powers reaching hundreds of kilowatts, and over milijoules of energy level. The main limitation to scaling of peak power and pulse energy in such systems is the onset of stimulated Brillouin scattering (SBS), which dynamics depends on pulse parameters.

In steady-state SBS (for pulse durations > 100 ns) the Brillouin gain is inversely proportional to the bandwidth. Therefore, the most effective approach to increase the SBS threshold involves broadening the spectral linewidth through external phase modulation \cite{Lago2011,Mu2018}. The modulation rate exceeds the inverse of the phonon lifetime, creating a multimode spectrum with a mode spacing larger than the gain linewidth. In this case, the coherence length of a laser source becomes much smaller than the interaction length, resulting in efficient SBS suppression. 
The interaction length can also be tailored by using a short fiber length with a large mode field area (LMA) to reach the desired power amplification level. The realization of this strategy allows the SBS threshold to increase significantly, reaching an average power level of hundreds watt for signle-frequency CW operation \cite{Liem:03, Jeong:05,500Wsingle}, and a nearly half milijoule of energy for pulsed operation within only 34 cm fiber length \cite{Fu2018}. An even more effective approach is not only to use the LMA fiber, but also to employ longitudinal variation of the large mode field diameter along the fiber length, the so-called Tapered Double-Clad Fibers (T-DCFs), representing a particularly effective implementation of this strategy. Their unique geometry, with a gradually increasing core diameter along the propagation axis, allows for a shift in the SBS gain spectrum and the evolving mode field area, which inherently raises the SBS threshold while enabling efficient high-power amplification within a compact system \cite{shiraki1995suppression, Huang2021, Patokoski:19, Lai:20}.

For transient SBS (for pulses <10 ns) the situation becomes more complicated. Despite the fact that the SBS threshold is still inversely proportional to the bandwidth and increases for shorter pulses \cite{Keaton:14}, the higher intensity pulses stimulate the Kerr nonlinear response, resulting in the growth of Stokes modes through space-time coupling \cite{Mauger_2010}. This process can be associated with other nonlinear effects such as transverse mode instability and, in the extreme cases, even self-focusing. In this scenario, the mitigation actions against SBS only do not work effectively, requiring the more complex strategy. The implementation of a few techniques simultaneously, such as spectral broadening, large mode area, and longitudinal variation of core diameter, result in successful peak power and energy scaling in the transient SBS regime with pulse energy up to 1.25~mJ~\cite{Scol2021} and peak powers as high as 180~kW~\cite{DiTeodoro2013, Scol2021}.  Table. \ref{tab:TDCF_SBS} summarizes key results from the literature that demonstrate a few nanosecond optical pulse amplification based on few mitigation approaches employed simultaneously by means of using PANDA T-DCFs as power amplifiers. The PT-DCF not only efficiently amplifies nanosecond signals, but also ensures the maintainance of the linear polarization state, which is an important parameter for many applications utilizing high-energy laser systems. However, independently of the SBS regime (steady state or transient) the linear polarized beam experienced a threshold about half lower for the SBS than the depolarized beam, limiting the energy/power scaling ~\cite{Stolen, Patokoski:19}. Although, SBS management realized through the random phase modulation technique enables dramatic progress in power/energy scaling, it inevitably degrades the temporal coherence of optical pulses~\cite{Mu2018, DiTeodoro2013, Scol2021}. For applications that require high spectral brightness, such as Doppler LIDAR or coherent lithography, this compromise is unacceptable, as minimal linewidth is critical for system performance.

\begin{table}[t]
\centering
\caption{Summary of reported nanosecond pulse MOPA systems using PT-DCF for SBS mitigation.}
\begin{tabular}{|l|c|c|c|c|c|c|}
\hline
\textbf{Ref.} 
& \makecell[c]{\textbf{Pulse energy} \\ (mJ)} 
& \makecell[c]{\textbf{Peak power} \\ (kW)} 
& \makecell[c]{\textbf{Pulse duration} \\ (ns)} 
& \makecell[c]{\textbf{Rep. rate} \\ (kHz)} 
& \makecell[c]{\textbf{Bandwidth}\\ (FWHM)} 
& \makecell[c]{\textbf{SBS mitigation measures} } \\
\hline
\cite{Belden2015}     & 0.019 & 8   & 2.5  & 400 & 0.7 GHz    & \makecell[c]{PT-DCF (60 µm core)} \\
\hline
\cite{Huang2021}      & 0.11  & 30  & 3.8  & 80  & 0.2838 GHz & \makecell[c]{PT-DCF (58 µm core)} \\
\hline
\cite{Mu2018}         & 1.0   & 100  & 10   & 10  & 40 GHz     & \makecell[c]{PT-DCF (60 µm core)\\spectral broadening} \\
\hline
\cite{DiTeodoro2013}  & 0.22  & 180 & 1.55 & 10  & 3.5 GHz    & \makecell[c]{PT-DCF (40 µm core)\\spectral broadening} \\
\hline
\cite{Scol2021}       & 1.25  & 170 & 3    & 5  & 167 pm     & \makecell[c]{PT-DCF (43 µm core)\\spectral broadening} \\
\hline
\end{tabular}
\label{tab:TDCF_SBS}
\end{table}

In this work, we present an approach for power/energy scaling in the transient SBS regime without external spectral modulation to preserve original coherence. SBS mitigation is achieved through the use of a spun ultra LMA T-DCF, known as an active spun T-DCF (sT-DCF), with an 80~\(\mu\)m core, bidirectionally pumped in the power amplification stage~\cite{Fedotov2021, Fedotov:21, Fathi:24}. This approach not only elevates the SBS threshold, but also mitigates polarization hole burning, enhances pump absorption, and enables efficient energy storage and amplification~\cite{Mazurczyk1994, Haar, Stolen, Patokoski:19, Fedotov:21, Fathi:24, Fathi2024-SR}.
The use of sT-DCFs maintains a stable state of polarization (SOP) by minimizing intrinsic birefringence via preform spinning during the fiber drawing process. This spinning uniformly distributes structural asymmetries and effectively cancels polarization errors accumulated during propagation. As a result, these fibers exhibit ultra-low birefringence (on the order of \(10^{-8}\)), rendering the SOP insensitive to thermal and mechanical perturbations. In this regard, sT-DCFs offer a robust alternative to existing techniques. The combination of tapering and spinning introduces a longitudinal variation of the mode field area and a distributed polarization response, both of which reduce the local optical intensity and shift the SBS gain spectrum along the fiber length. Therefore, sT-DCFs inherently suppress SBS without spectral line broadening and simultaneously maintain the linear polarization state, making them ideally suited for power/energy scaling of narrow-linewidth nanosecond MOPA systems. In this paper, we experimentally demonstrate a narrow linewidth (53.8 MHz), 8 ns MOPA system operating at 100 kHz, delivering pulses with 1.6 mJ energy, 160 W average power, and 188 kW peak power. The output exhibits near-diffraction-limited beam quality (\( M^2 \sim 1.3 \)) and a highly stable linear polarization state with a degree of polarization of 97.5\%.

\section*{Experimental setup}

\begin{figure*}[t]
\centering
\includegraphics[width=\linewidth]{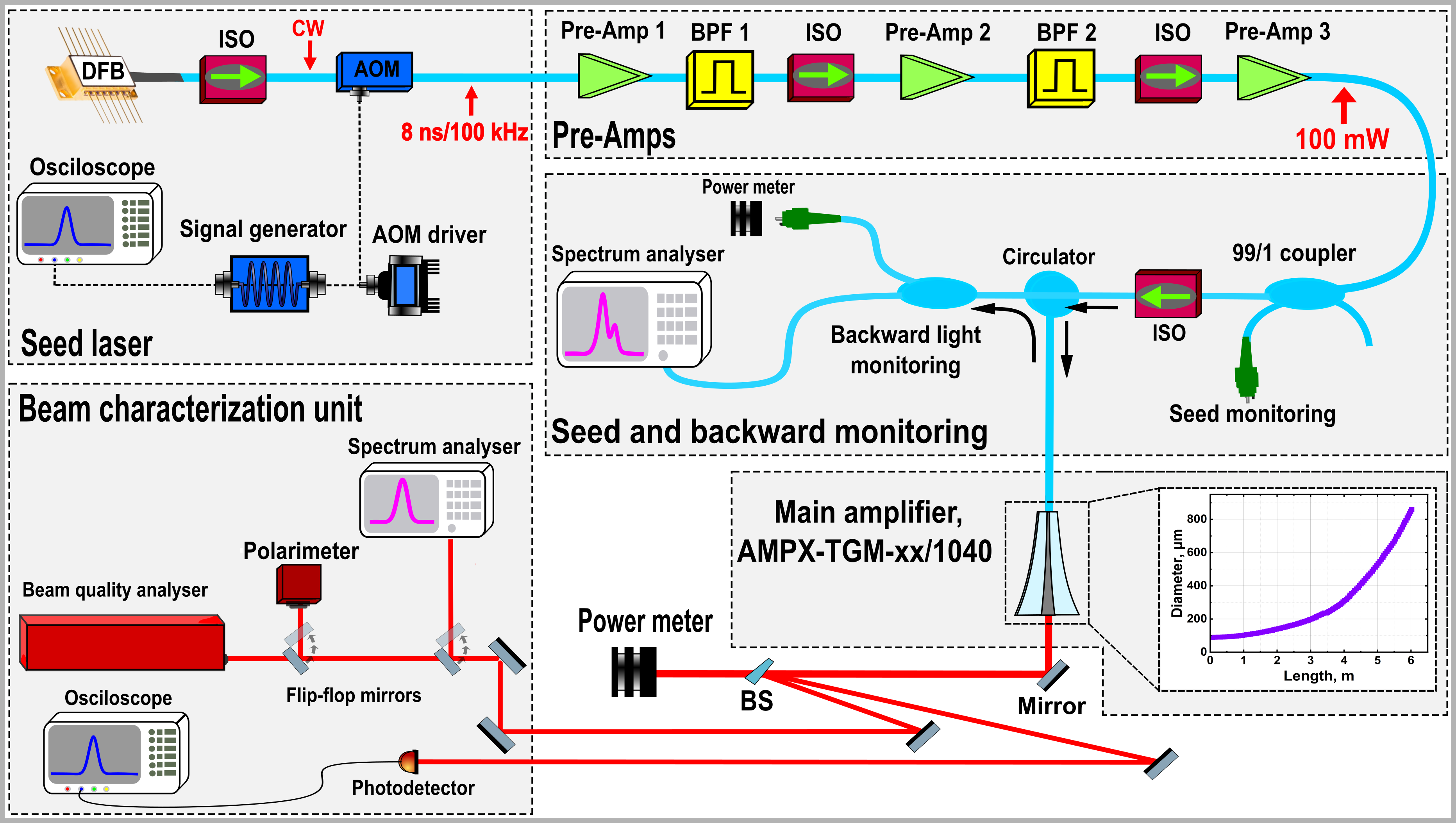}
\caption{Schematic representation of the experimental setup based on master oscillator-power amplifier (MOPA) architecture. The components are labeled as follows: DFB, semiconductor distributed feedback; ISO, Isolator; AOM, acousto-optic modulator; Amp, amplifier; BPF, bandpass filter; and BS, beam sampler. The inset is a sT-DCF longitudinal profile.}
\label{setup-high energy}
\end{figure*}

The laser system is built on the basis of the master-oscillator-power amplifier (MOPA) architecture. Fig. \ref{setup-high energy} illustrates an MOPA scheme with three main parts: a front-end (F-E) seed laser system, a main amplifier and the beam characterization unit. The front-end laser source consists of a semiconductor distributed-feedback (DFB) laser operating at 1040 nm in continuous-wave (CW) mode, signal modulation stage, and a chain of three pre-amplifiers. The DFB laser has a spectral linewidth of 10 MHz measured using the self-heterodyne technique.
The CW output of the DFB is then externally modulated by an acousto-optic modulator (AOM) to generate 8-ns optical pulses at a 100 kHz repetition rate.
An isolator is placed before the AOM to prevent unwanted back reflections into the DFB laser. The optical pulse is then amplified through two successive stages of the pre-amplifier. Each pre-amplifier stage consists of 60 cm length Ytterbium-doped fiber (PM401), pumped by a single-mode laser diode operating at 976 nm. After each amplifier, a narrow bandpass filter (BPF) of 2 nm filters out the signal from amplified spontaneous emission (ASE), and an isolator blocks any backward ASE from the subsequent stage. 
 At the third stage of the pre-amplifier the signal is amplified up to 100~mW using 1.8~m of ytterbium-doped fiber. A following 99:1 optical splitter is used to monitor the seed output. The 99\% port directs the signal to the main amplifier based on sT-DCF. To monitor any backward-propagating light that returns from the main amplifier toward the F-E seed system, an optical circulator is placed between the seed and the main amplifier.

The main amplifier stage employs an sT-DCF that is bidirectionally pumped. The spin pitch is set to 50~mm, and the fiber geometry, including the core/cladding dimensions and the taper length, has been optimized to enhance pump absorption while maintaining the mode content. The active core is fabricated using the modified chemical vapor deposition (MCVD) technique, resulting in numerical apertures (NA) of 0.08/0.27/0.48 for the core, the first cladding and the second cladding, respectively. The fiber provides a high core absorption of 800~dB/m at 976~nm. The taper length is approximately 6~m, and the diameters transit from 8.3/75/90~µm to 78/750/850~µm for the core, first cladding, and second cladding, respectively. The longitudinal geometry of the sT-DCF is illustrated in the inset of Fig.~\ref{setup-high energy}.
The pump architecture consists of seven wavelength-stabilized laser diodes at 976~nm, each delivering up to 120~W, combining for a total pump power exceeding 840~W. These are merged using a 7×1 fused-fiber pump combiner that couples the pump light into a single fiber with a 200~µm core diameter. This output is directly spliced to the wide end of the sT-DCF in a backward-pumping configuration, enabling alignment-free integration and improving overall system compactness. In addition, 18~W of pump power at 915~nm is injected into the narrow side of the taper using a (2+1)×1 combiner. This dual-sided pumping scheme ensures uniform gain distribution and efficient energy transfer along the full taper length, supporting the amplification of high-energy nanosecond pulses while minimizing nonlinear effects and maintaining a compact amplifier design.

An uncoated fused silica wedged window is used as a beam sampler to direct a portion of the beam to the characterization unit, enabling measurement of key output properties such as the optical spectrum, beam quality factor ($M^2$), polarization, pulse shape and train, and spatial and temporal coherence.

\section*{Experimental results}
\begin{figure}[t]
\centering
\begin{subfigure}{.5\textwidth}
  \centering
  \includegraphics[width=0.94\linewidth]{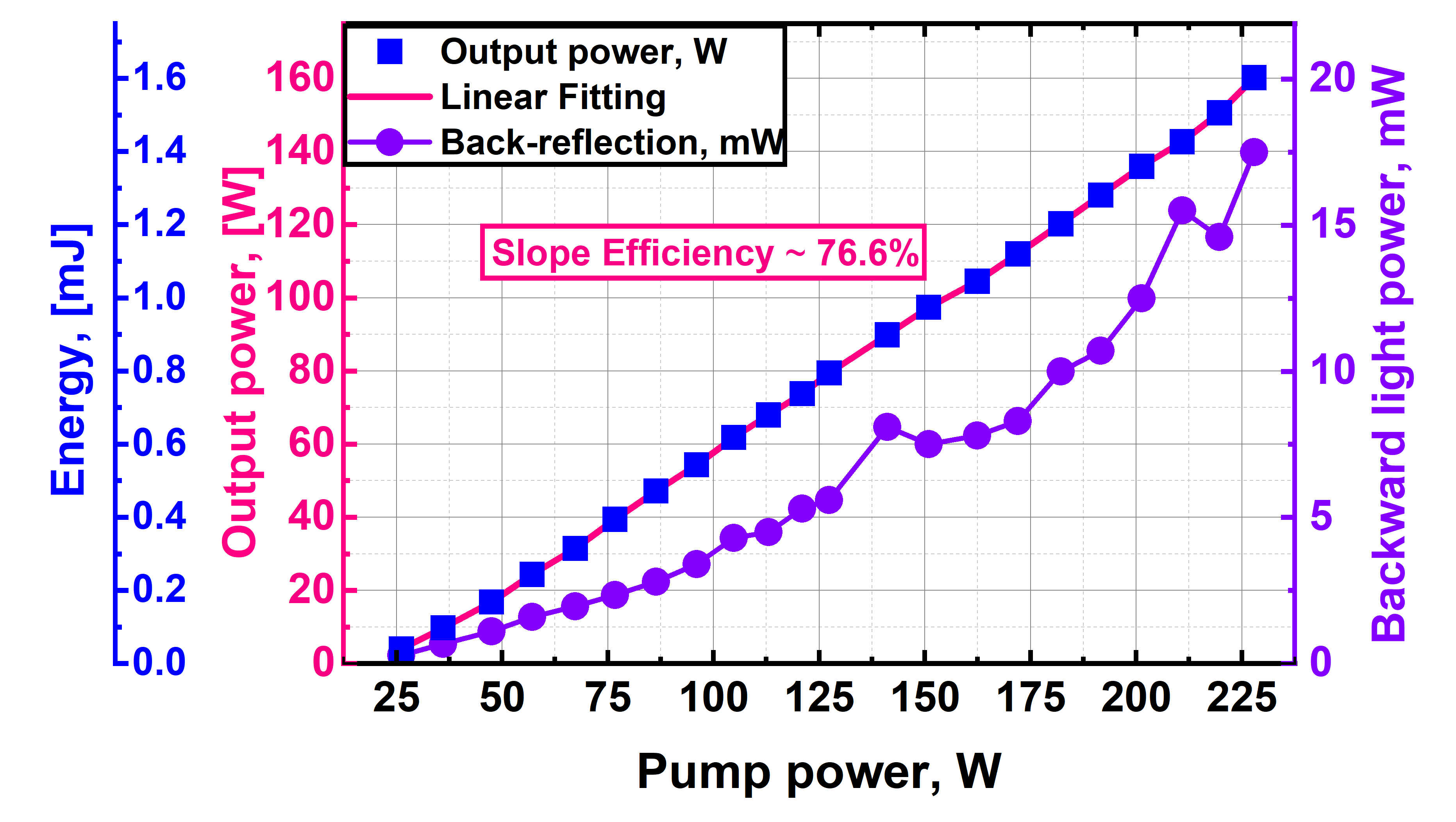}
  \captionsetup{justification=centering}
  \caption{}
  \label{Slope}
\end{subfigure}%
\begin{subfigure}{.45\textwidth}
  \centering
  \includegraphics[width=0.92\linewidth]{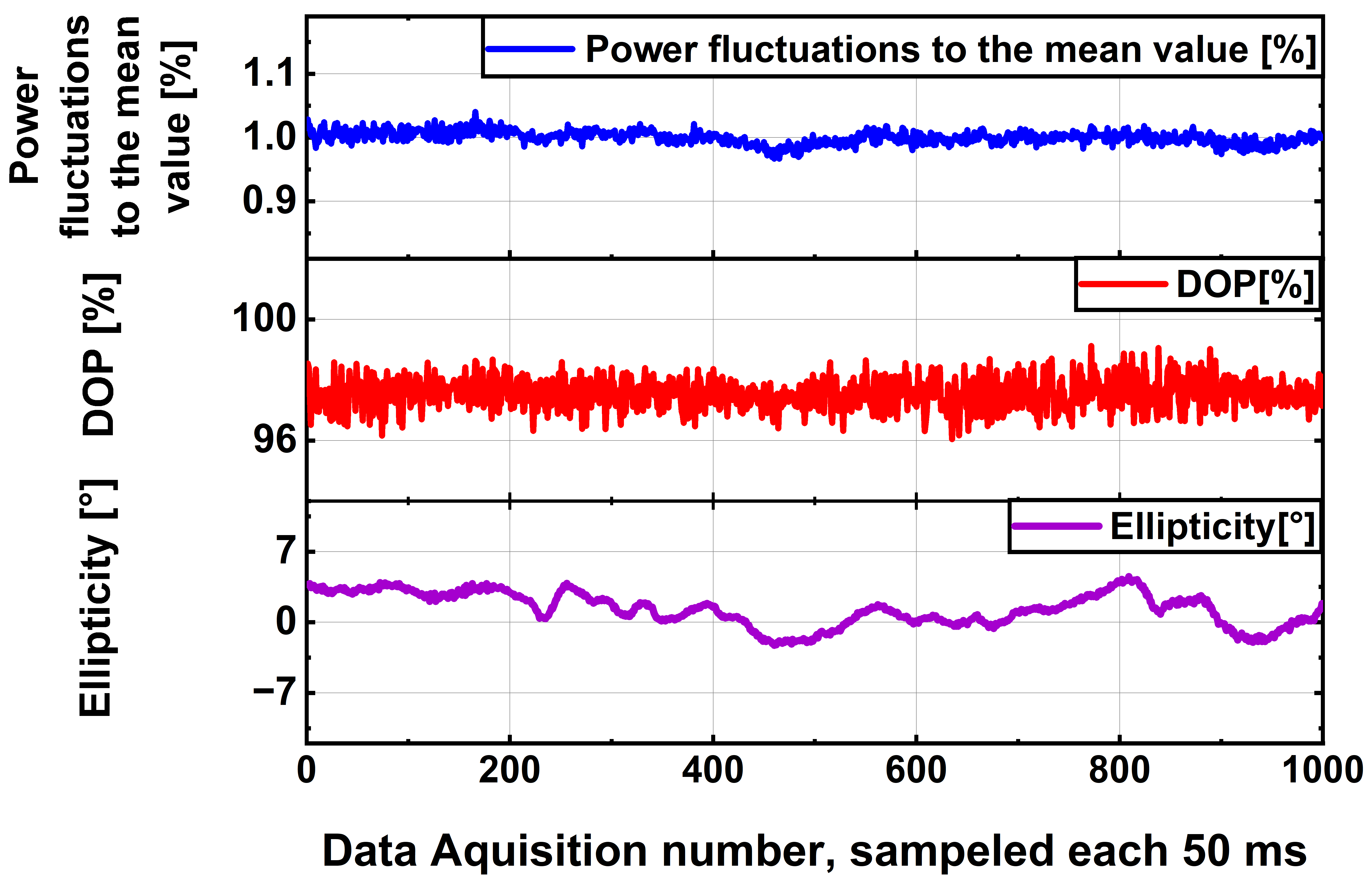}
  \captionsetup{justification=centering}
  \caption{}
  \label{Pol}
\end{subfigure}
\caption{(a) The laser output power/energy and the backward-propagating light power of the 1.6 mJ / 8 ns laser system as a function of total launched pump power, illustrating power/energy scaling performance; (b) stability analysis of output power and polarization. A total of 1000 samples were acquired at 50 ms intervals. Output power stability is shown as the normalized signal relative to its mean value.}
\label{power and Pol}
\end{figure}

Fig.~\ref{Slope} shows the output power/energy of the laser system and the backward-propagating light power versus the total launched pump power. The sT-DCF amplifier delivers up to 160~W of average power ($\simeq 1.6~\mathrm{mJ}$ of pulse energy), achieving a slope efficiency of 76.6\%. The backward-propagating light power was measured to be below 20~mW at the highest energy level. At maximum output (1.6~mJ), the system maintains a high polarization quality, demonstrating an average degree of polarization (DOP) exceeding 97.5\%, as shown in Fig.~\ref{Pol}.
The standard deviations of the output power and the DOP of the beam were calculated as < 1\% and 0.5\%, respectively. The ellipticity of the beam did not exceed 4.17 degrees and varies within a standard deviation of 1.46 degrees. Polarization was measured over 1000 samples with a 50 second interval using a
commercial polarimeter (PAX1000IR2/M).

\begin{figure}[ht]
\centering
\begin{subfigure}{.46\textwidth}
  \centering
  \includegraphics[width=.8\linewidth]{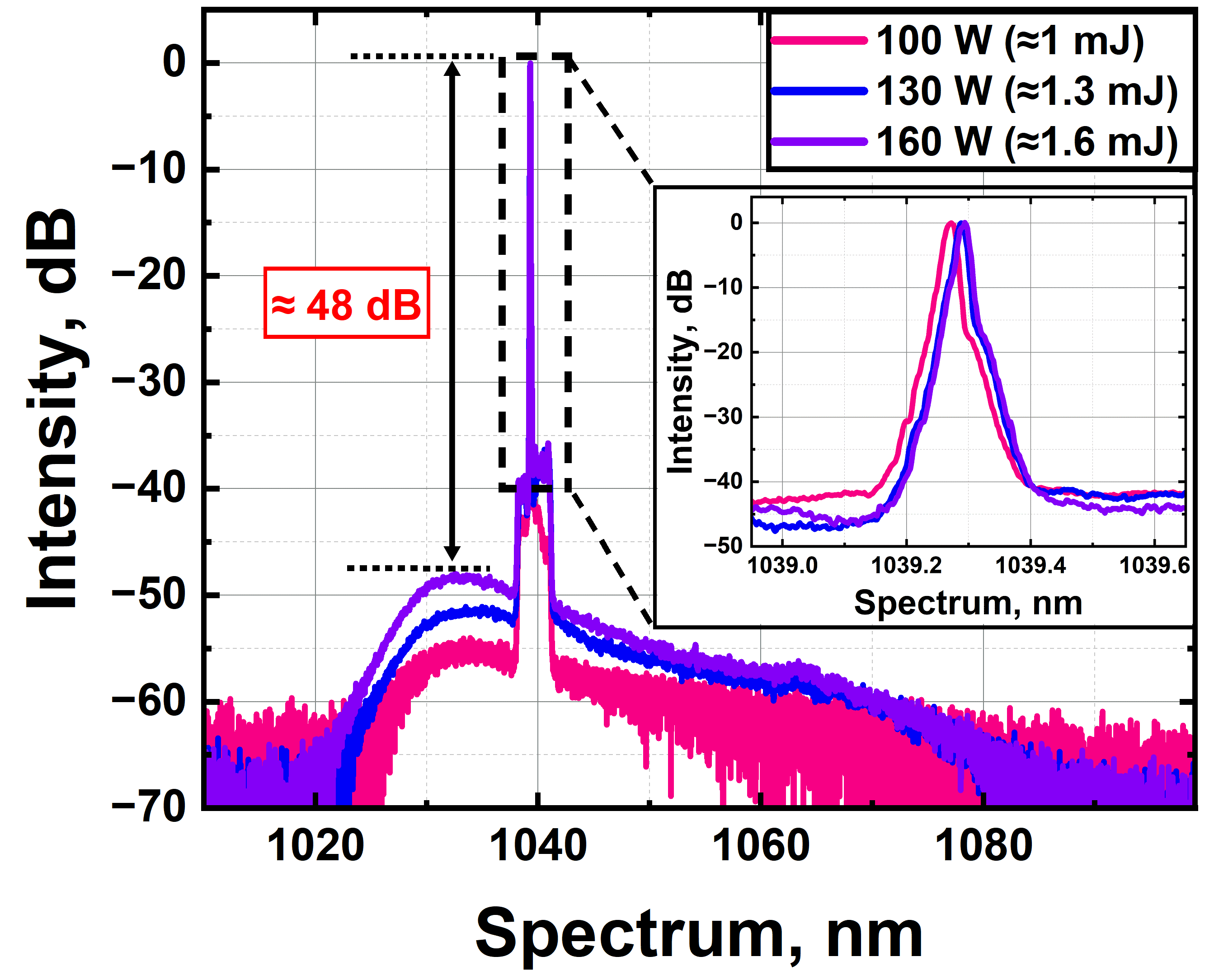}
  \captionsetup{justification=centering}
  \caption{}
  \label{Spec-wide}
\end{subfigure}%
\begin{subfigure}{.46\textwidth}
  \centering
  \includegraphics[width=.81\linewidth]{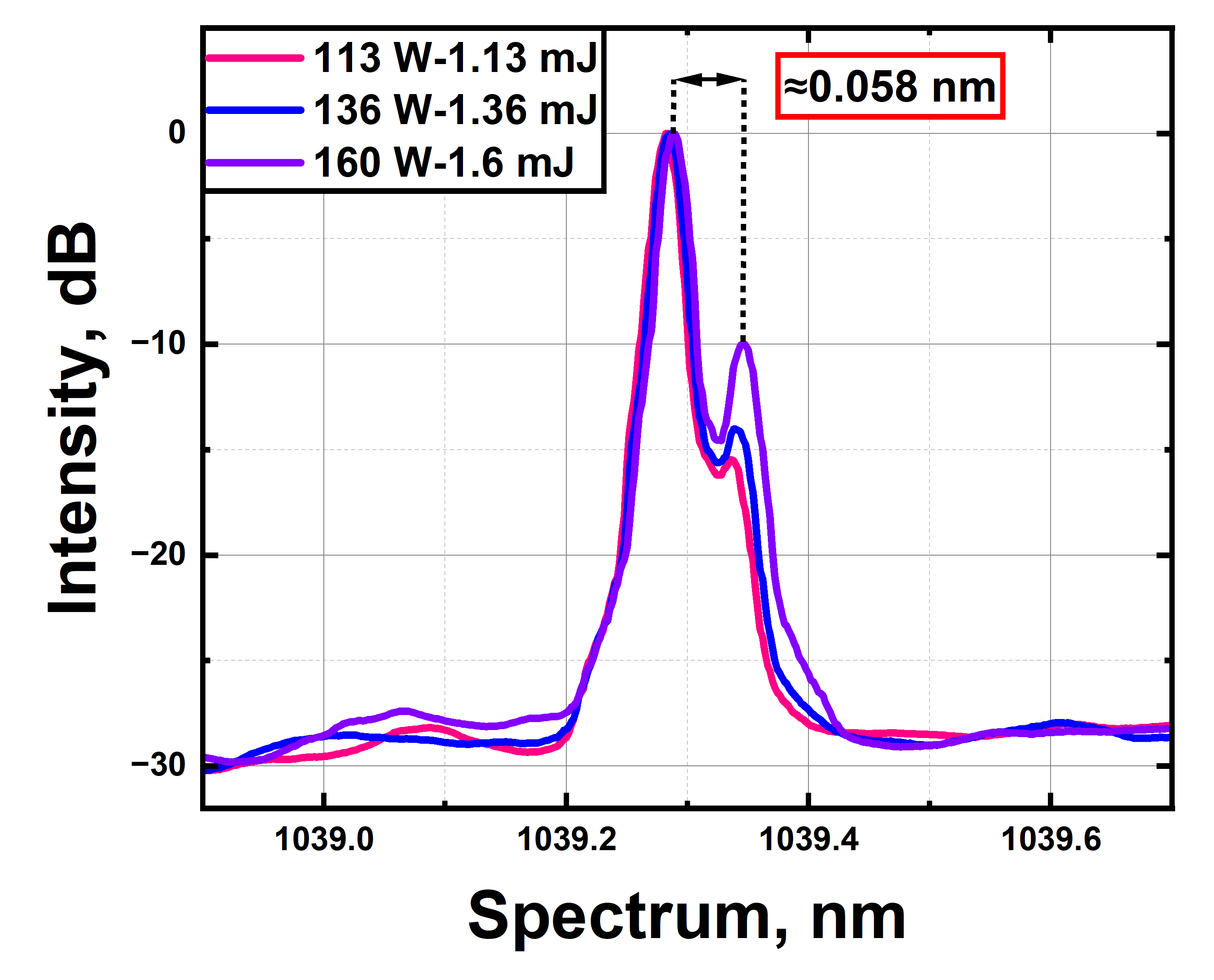}
  \captionsetup{justification=centering}
  \caption{}
   \label{Backward}
\end{subfigure}
\caption{Optical spectrum characterization of the amplified signal at different output powers/energy levels. (a) Optical spectrum of the amplified signal over a wide spectral range. The inset shows the optical spectrum of the amplified (forward) signal over a narrow spectral range; and (b) Optical spectrum of the backward-propagating light in the narrow spectral range.}
\label{Spectrum}
\end{figure}

The optical spectra of the amplified signals across a wide wavelength range for different power/energy levels
are presented in Fig.~\ref{Spec-wide}. The spectra reveal a low ASE background and the absence of the stimulated Raman scattering (SRS) growth, indicating efficient direct amplification of nanosecond pulses. The inset shows the optical spectra of the amplified signals in the narrow wavelength range for different power/energy levels. 
Fig.~\ref{Backward} presents the optical spectrum of backward propagating light collected from the circulator port in high-energy operation. A distinct spectral peak appears at approximately +0.058~nm relative to the central signal wavelength. This frequency-shifted component is a clear indication of SBS, originating from the interaction between the optical signal and thermally induced acoustic phonons within the fiber.
The measured Stokes shift, 0.058~nm / 16.1~GHz, corresponds well to the expected SBS frequency shift in silica fibers at a center wavelength of 1040~nm.
At the maximum output power/energy, the signal-to-SBS peak ratio is measured to be approximately 10~dB, suggesting that although SBS is initiated, it remains well suppressed relative to the main signal. The backward light reaches a maximum 17 mW average power, which reveals the effectiveness of SBS mitigation strategies in sT-DCF-based high-power fiber amplifier systems.

\begin{figure}[ht]
\centering
\begin{subfigure}{.46\textwidth}
  \centering
  \includegraphics[width=0.8\linewidth]{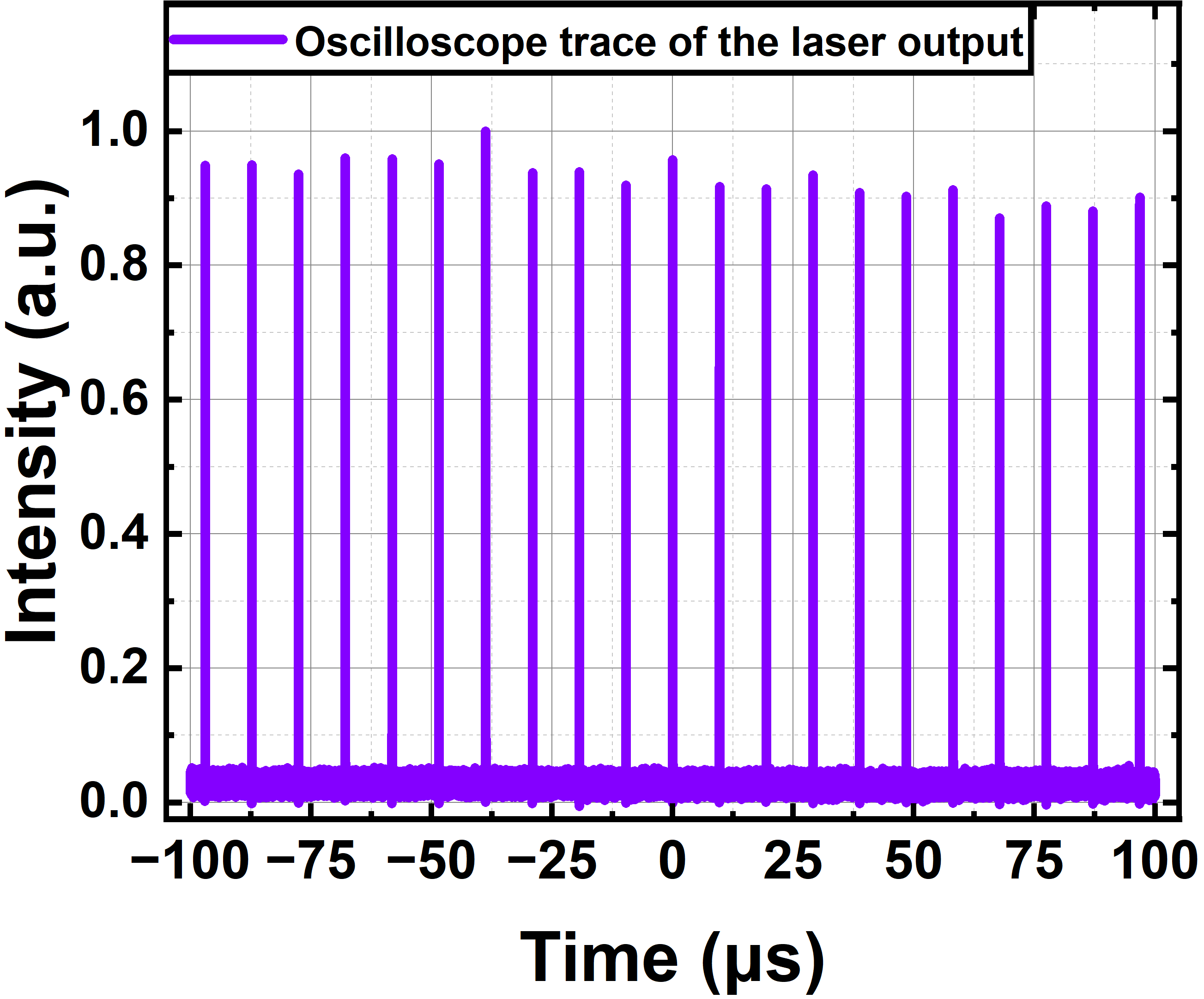}
  \captionsetup{justification=centering}
  \caption{}
  \label{Pulse-trace}
\end{subfigure}%
\begin{subfigure}{.46\textwidth}
  \centering
  \includegraphics[width=0.82\linewidth]{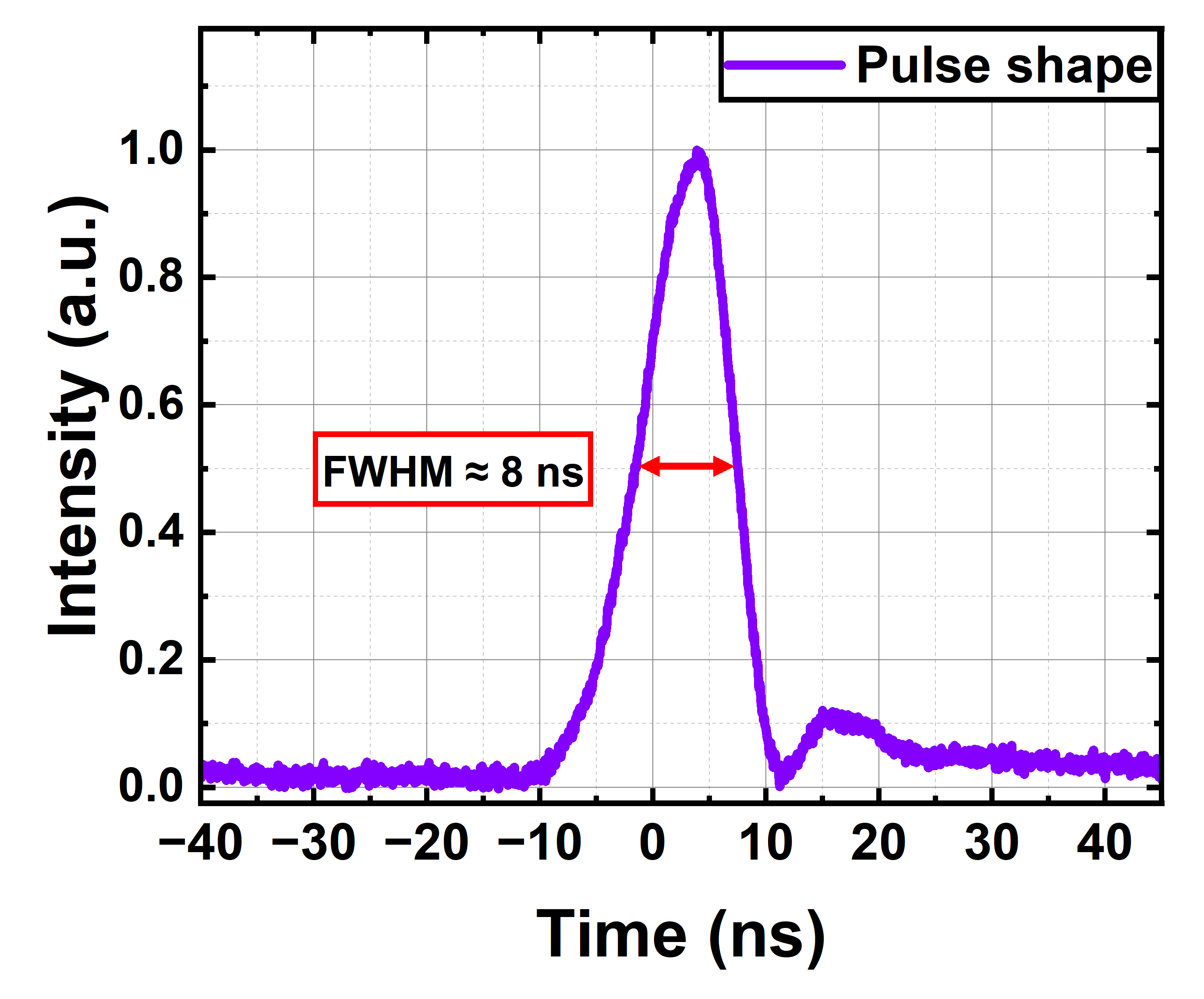}
  \captionsetup{justification=centering}
  \caption{}
  \label{Pulse-shape}
\end{subfigure}
\caption{Oscilloscope traces of the laser output operating at a 100 kHz repetition rate. (a) Pulse train (25 µs/div); (b) Single pulse profile (10 ns/div), showing a pulse shape with an 8 ns FWHM.}
\label{Pulse trace and shape}
\end{figure}

The temporal characteristics of the laser output were characterized by measuring the pulse train and individual pulse profile using a 1.2 GHz bandwidth photodetector and a 2.5 GHz bandwidth digital oscilloscope, as shown in Fig.~\ref{Pulse trace and shape}.
The laser operates at a repetition rate of 100 kHz, corresponding to a 10 µs interval between successive pulses. The pulse train (Fig.~\ref{Pulse-trace}) confirms stable and periodic output at the expected repetition rate. A zoomed-in view of a single pulse (Fig.~\ref{Pulse-shape}) reveals a nearly Gaussian temporal profile with an FWHM of approximately 8 ns, indicating clean temporal behavior without significant pre- or post-pulse structures.

\begin{figure}[ht]
\centering
\begin{subfigure}{.55\textwidth}
  \centering
  \includegraphics[width=0.94\linewidth]{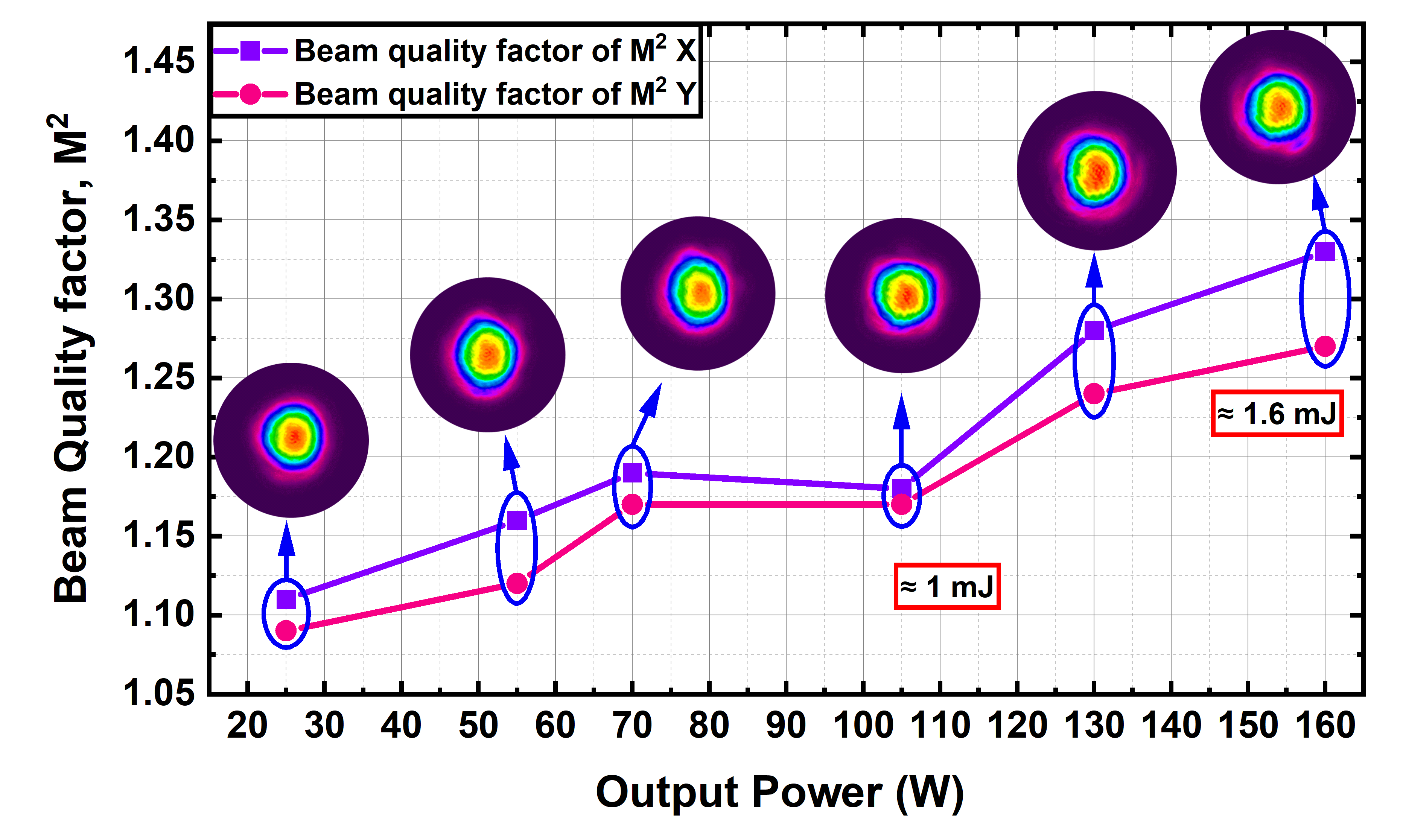}
  \captionsetup{justification=centering}
  \caption{}
  \label{M2-ev}
\end{subfigure}%
\begin{subfigure}{.43\textwidth}
  \centering
  \includegraphics[width=0.92\linewidth]{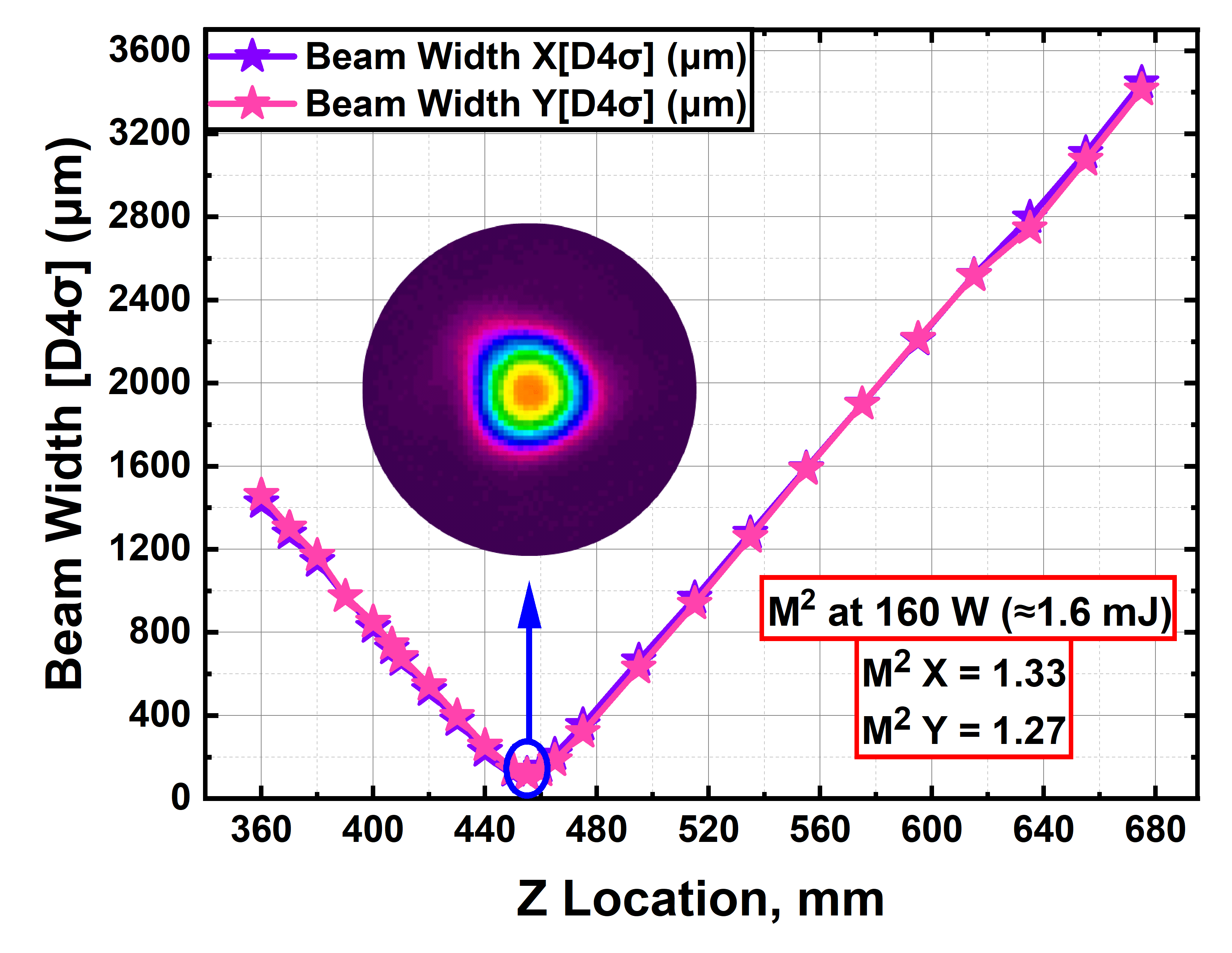}
  \captionsetup{justification=centering}
  \caption{}
  \label{M2-1.6mj}
\end{subfigure}
\caption{Beam quality characterization of the amplified signal. (a) Measured \(M^2\) values at different output power levels, shown alongside corresponding near-field beam profiles; (b) \(M^2\) curve of the amplified beam at 1.6 mJ pulse energy (160 W average power);
the inset shows the beam profile at the focal plane of the $M^2$ meter.}
\label{Beam quality}
\end{figure}

Fig.~\ref{Beam quality} presents the beam quality characterization of the amplified signal based on the $M^2$ factor measurements. The evolution of beam quality with increasing output power, along with the corresponding near-field beam profiles, is shown in Fig.~\ref{M2-ev}. At maximum average output power and pulse energy (160~W / 1.6~mJ), the beam quality was measured to be $M^2_X = 1.33$ and $M^2_Y = 1.27$, indicating no signs of mode instability (see Fig.~\ref{M2-1.6mj}; the inset shows the beam profile in the focal plane of the $M^2$ meter).

\subsection*{Coherence assessment}

As mentioned in the description of the experimental setup, the linewidth of the DFB seed laser in continuous wave operation (before AOM) is approximately 10 MHz. Since temporal coherence and the spectrum are connected by a Fourier transform, this corresponds to a temporal coherence length of about 9.5 meters under the 1/$e$ definition. The AOM introduces the modulation of the signal forming pulses of 8 ns. This unavoidably causes the change in the spectrum and coherence length. After the AOM, the pulse length is approximately 8 ns, which corresponds to 2.4 meters of coherence length. As the coherence length cannot be greater than the pulse length, therefore, the AOM must cause a change in the spectrum of the seed laser. Following this, the signal undergoes several stages of amplification, which may additionally introduce the degradation of coherence via spectrum broadening or pulse distortion. We perform the measurement of both spatial and temporal coherence at the output of F-E seed and the main amplifier with the wavefront-folding interferometer (WFI), which has been proven to be an efficient and fast device for coherence measurements; see Refs.~\cite{Koivurova:19,Halder:20,Turunen:22} for more information. 

\begin{figure}
    \centering
    \includegraphics[width=0.9\linewidth]{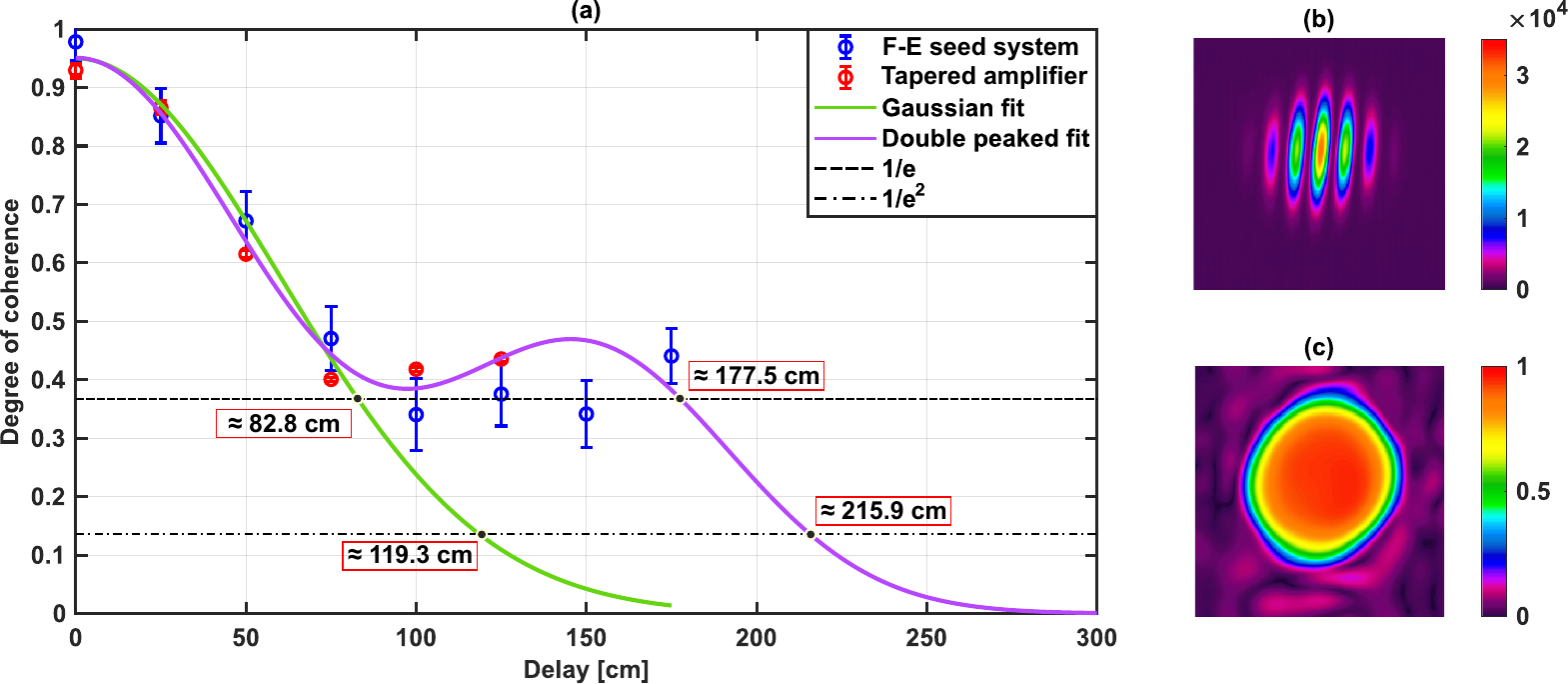}
    \caption{(a) Temporal coherence of the seed and amplified pulse with a Gaussian fit, (b) typical interference fringes in a spatial coherence measurement, and (c) the degree of spatial coherence across the beam at the output of the main amplifier.}
    \label{fig:coherence}
\end{figure}

Fig.~\ref{fig:coherence} features the measured coherence properties. In panel (a), we present the results of the temporal coherence measurement. The blue data points correspond to the pulse train after the F-E seed system, while the red is after the main amplifier (10 W average power). The circles are the mean values of the temporal degree of coherence as calculated over the whole spatial cross-section of the beam at each delay, while the error bars represent the standard deviation from the mean. Note that the values measured after the F-E seed system featured a significant amount of random variance. After the main amplifier, the temporal coherence curve remained about the same shape, which is expected from similar shapes of the spectra presented in Fig.~\ref{Spectrum}. However, the amount of random noise was sharply decreased, and thus the error bars after amplification are much smaller. Due to technical difficulties, we were unable to measure the temporal coherence after the tapered amplifier for delays larger than 125 cm. We fit a Gaussian curve to the measured temporal coherence data with a 1/$e$ width of about 82.8 cm ($\sim$115.2 MHz in spectral bandwidth). Note, however, that a Gaussian is not representative of the true temporal coherence curve but rather presents a lower bound on the coherence length (upper bound on spectral bandwidth). As an additional constraint, we fit a double-peaked Gaussian distribution on to the data points. For this distribution, we emphasized the weight of the pre-amplified data-points, since they have a lower uncertainty. The 1/$e$ width of this distribution is about 177.5 cm ($\sim$53.8 MHz). Moreover, the distribution fits well within the measured pulse width of about 8 ns ($\sim$240 cm, or $\sim$40 MHz).

Having discussed the beam quality as in Fig.~\ref{Beam quality}, it is important to point out it's relationship with spatial coherence. For a partially coherent Gaussian-Schell model beam, the beam quality is inversely proportional to the overall degree of spatial coherence $\bar\gamma$, via the relation $M^2 = 1/\bar\gamma^2$ \cite{Pu:98,Lajunen:04}. In other words, if the beam has a perfect Gaussian cross-section, then an $M^2$ of the order of 1.3 would correspond to an overall degree of coherence of about $\bar\gamma=0.88$. Since the cross-section of the beam is clearly not Gaussian, the quality measurements presented in Fig.~\ref{Beam quality} already indicate a high degree of spatial coherence. That is, the slight increase in $M^2$ is likely due to non-Gaussian intensity distribution. In Fig.~\ref{fig:coherence}(b) we present a depiction of a typical interference pattern in spatial coherence measurements, and as one can see the visibility is high throughout the beam cross-section. In panel (c), we further show the corresponding degree of spatial coherence, which has a value of $\gamma = 0.94 \pm 0.01$ in high intensity areas.

\section*{Conclusion and discussion}

In this work, we have demonstrated a high-energy, narrow-linewidth, nanosecond fiber MOPA laser system based on an all-glass, sT-DCF amplifier, achieving 1.6~mJ pulse energy with 8~ns pulse duration at a repetition rate of 100~kHz. The system delivers pulses with 53.8~MHz linewidth, a beam quality factor of \(M^2 \sim 1.3\), slope efficiency of 76.6\%, and a degree of polarization exceeding 97.5\%. A high degree of spatial coherence ($\gamma = 0.94 $) was also maintained throughout the entire power range.

Unlike other state-of-the-art systems that rely on phase modulation to suppress SBS, our approach uses the inherent properties of sT-DCF amplifiers, gradual mode field expansion, and distributed polarization response to simultaneously suppress SBS, stabilize polarization, and maintain high coherence. The absence of spectral broadening allows us to preserve the narrow linewidth, making the system suitable for applications where spectral purity is paramount, such as Doppler LIDAR, and coherent beam combining.
Our results compare favorably with previous works that employed PANDA-type T-DCFs in combination with external SBS suppression techniques such as spectral broadening~\cite{Mu2018, Scol2021, DiTeodoro2013}. Although these systems achieved high pulse energies, they often did so at the expense of temporal coherence, with optical linewidths typically exceeding several hundred MHz. In contrast, our amplifier maintains a sub-100~MHz linewidth without any spectral modulation, while delivering higher pulse energy than previously reported systems relying solely on passive SBS mitigation using PT-DCFs~\cite{Belden2015, Huang2021}.

This work highlights the effectiveness of sT-DCFs as a robust and alignment-free platform for building compact, high-performance fiber-based MOPA systems. By avoiding free-space components and additional SBS suppression elements, our approach reduces complexity and increases system stability, enabling practical deployment in field applications. Future work will focus on scaling this system to even higher pulse energies, optimizing fiber geometry for further SBS suppression.


\section*{Acknowledgements}

The authors thank Andrey Chumachenko, Konstantin Miroshnichenko, and Andrei Gurovich for their valuable contributions to this work.

\section*{Funding.}
Part of this work has been supported by the European Commission Horizon2020 program (HoLiSTEP  project- grant agreement Nr. 101137624) and the Flagship for Photonics Research and Innovation (PREIN). 

\section*{Author contributions}
H.F., R.G., and V.F. conceived the experiments. H.F., U.S., E.M., and E.A. conducted the experiments. H.F. analyzed the results. M.K. carried out the theoretical and simulation analysis of the coherence properties. A.G. supervised manufacturing of the main tapered amplifier.  H.F. wrote the manuscript. All authors reviewed and approved the final manuscript.

\section*{Competing interests}
The authors declare no conflicts of interest.

\section*{Data availability statement.}
The datasets used and/or analysed during the current study are available from the corresponding author on reasonable request.

\section*{Additional information}
\textbf{Correspondence} and requests for materials should be addressed to H.F.

\end{document}